\documentstyle[aps,preprint]{revtex}
\tightenlines
\begin{document}
\draft
\title{Gauge string in the fermion asymmetric matter.}
\author{A.A. Kozhevnikov}
\address{Laboratory of Theoretical Physics,\\
S.L. Sobolev Institute for Mathematics,\\
630090, Novosibirsk-90, Russian Federation
\footnote{Electronic address: kozhev@math.nsc.ru}
}
\date{\today}
\maketitle
\begin{abstract}
Two new effects of interaction of the gauge string  with a
homogeneous density of fermions are considered in a gauge model
with an anomalous coupling of vector fields with fermions. First, the
presence of an induced nonzero magnetic-like
helicity on the straight string is
demonstrated. Second, it is shown that the equation of motion of the string
is modified by a nonlinear term that can be decomposed into the
correction to the string tension and  an additional force
perpendicular to the tangent and normal vectors of the string.
Static configurations are found and their stability is studied.
\end{abstract}


\narrowtext

The presence of a large asymmetry of some fermionic charges, in particular,
the lepton number, constitutes an interesting theoretical possibility
\cite{linde,riotto98,shaposh}. Current experimental bounds on this type of
asymmetry are  large enough \cite{bounds}. On the other hand,
the stringlike defects analogous to those already observed in type II
superconductors immersed into magnetic field, might have been produced
at early epoch of the evolution of the Universe
\cite{vilenkin94,kibble95}. A wide class of such defects is described as
a classical solution to some gauge field theoretical models like an Abelian
Higgs model (AHM) \cite{ano}. In particular, the Z string  \cite{z}
is an embedding of the Abrikosov-Nielsen-Olesen solution \cite{ano}
into SU(2)$\times$U(1) electroweak model. It is important that the gauge
field configuration of a Z string contributes \cite{vachas94} to the right
hand side of the anomaly equation  for the sum of the baryon and lepton
numbers. The above developments raise the question of could a nonzero asymmetry
of some fermionic charge  exert an influence on the  string motion?

Since the pioneering work \cite{forst} devoted to the derivation of the
string equation of motion from AHM in the zero thickness limit,
attempts were undertaken to obtain the finite
thickness corrections to the string motion \cite{orland,arodz,bonjour}.
Here we will find the correction of basically different
origin which arises due to the presence of a nonzero density of some fermionic
charge characterized by a nonzero chemical potential, provided an underlying
gauge model contains an anomalous interaction of gauge fields with chiral
fermions. The latter takes place, for example, in the case of a Z string and
the lepton and baryon numbers of the chiral leptons and quarks of the
electroweak theory. The fact is that the action of the model should then
contain an additional term
\begin{equation}
\Delta S=\int dt\mu n_{\rm CS},
\label{delac}
\end{equation}
\cite{redlich85,rubakov85},
where $\mu$ is the chemical potential of fermions,
and $n_{\rm CS}$ is the Chern-Simons number of the gauge field.
Hereafter the case of zero temperature is considered, $T=0$.
Since some parity-non-invariant configurations of the
gauge field trapped by the string are known to have a nonzero $n_{\rm CS}$
\cite{vachas94,sato95,kozhev95}, the above term in the action should
manifest itself in the string equation of motion. Below corresponding
correction is found explicitly. As a by-product, the magnetic-like helicity
induced on the straight gauge string  by a nonzero fermionic density is
evaluated.

In order the presentation to be self-contained, let us sketch the derivation
of the string equation of motion from the action of AHM.
The gauge coupling constant $g$, the self-coupling of the Higgs field
$\lambda^2$, and its vacuum expectation value $\eta^2/2$ are such that
$m_H\gg m_V$; $\ln m_H/m_V$ is also large, where
$m_V=g\eta$ and $m_H=\lambda\eta$ are the
masses of the gauge and Higgs bosons, respectively. This is the London limit
of AHM. In this limit, the Higgs field is $\phi=\eta\exp i\chi/\sqrt{2}$
everywhere except the
string line where $|\phi|$ vanishes at the distance scale
$\sim m^{-1}_H$, and $\chi$ is the spacetime dependent phase. 
Then the AHM action can be written as
\begin{equation}
S=\int d^4x\left[-F^2_{\mu\nu}/4+
m^2_V\left(A_\mu+\partial_\mu\chi/g\right)^2/2\right],
\label{actill}
\end{equation}
where $A_\mu=(A_0,{\bf A})$,
$F_{\mu\nu}=\partial_\mu A_\nu-\partial_\nu A_\mu$ being the vector potential
and the field strength tensor, respectively.
The contribution of the Higgs field potential is suppressed in this limit
by the factor $1/\ln m_H/m_V\ll1$ as compared to the kinetic energy of this
field, see below, and hence is neglected.
Note that the case of Z-string is obtained by the replacement
$A_\mu\to Z_\mu$, $g\to\sqrt{g^2+g^{\prime 2}}/2$, where $Z_\mu$ is the
vector potential of Z boson, $g$ and $g^\prime$
are the SU(2) and U(1) coupling constants of the  electroweak theory.
The field equation for the vector potential obtained from the variation of
the action (\ref{actill}) under the condition $\partial_\mu A^\mu=0$, is
$(\partial^2+m^2_V)A_\mu=-m^2_V\partial_\mu\chi/g$.
It is solved to give
\begin{eqnarray}
A_\mu+\partial_\mu\chi/g&=&
\left[1-m^2_V(\partial^2+m^2_V)^{-1}_{\rm ret}\right]
\partial_\mu\chi/g, \nonumber\\
F_{\mu\nu}&=&-m^2_V(\partial^2+m^2_V)^{-1}
_{\rm ret}\partial_{[\mu}\partial_{\mu]}\chi,
\label{amu}
\end{eqnarray}
where the subscript points to the choice of the Green function as the retarded
one. The expression for the commutator of derivatives of the singular
phase of the Higgs field  is \cite{orland}
\begin{equation}
\partial_{[\mu}\partial_{\mu]}\chi=2\pi\varepsilon_{\mu\nu\alpha\beta}
\int d^2s\partial_\tau X^\alpha\partial_\sigma X^\beta
\delta^{(4)}(x-X),
\label{chi}
\end{equation}
where $X\equiv X(\sigma,\tau)$ is the string world sheet,
$s^A=(\tau,\sigma)$ is the two-vector, and the unit winding
number of the Higgs field is taken.
In what follows, the space Fourier transforms of the gauge field configuration
found earlier \cite{kozhev99} is often used.
This form is especially comfortable
because the fields are expressed through elementary functions rather than the
special ones. Using (\ref{chi}) one can obtain the explicit form for
the field strength tensor, the gauge invariant combination of the vector
potential, and the vector $v_\mu\equiv\partial_\mu\chi/g$ as
\begin{eqnarray}
F_{\mu\nu}(k)&=&-{2\pi\over g}\varepsilon_{\mu\nu\alpha\beta}
\int d^2s\partial_\tau X^\alpha\partial_\sigma X^\beta
{m^2_V\over m^2_V-k^2-i\varepsilon k^0}\exp(ikX),   \nonumber\\
\left(A_\mu+\partial_\mu\chi/g\right)(k)&=&
-{k^2\over m^2_V-k^2-i\varepsilon k^0}v_\mu(k),  \nonumber\\
v_\mu(k)&=&-{2\pi i\over g}\varepsilon_{\mu\nu\alpha\beta}
{k^\nu\over k^2}\int d^2s\partial_\tau X^\alpha\partial_\sigma X^\beta
\exp(ikX).
\label{fieldsk}
\end{eqnarray}
In the case of many contours one should sum over all of them.
The insertion of Eq. (\ref{fieldsk}) into Eq. (\ref{actill})
gives the gauge vortex string action in the London limit.
One can
observe that, similar to the potential energy of the Higgs field,
the contribution of the gauge field strength to the action is suppressed,
in this limit, by
the same factor $1/\ln m_H/m_V\ll1$. This is due to the $k^\nu$ factor in the
numerator of the expression for the gauge invariant combination of the vector
potential. Then the dominant contribution to the action becomes 
\begin{eqnarray}
S&=&-{m^2_V\over2}\left({2\pi\over g}\right)^2\int{d^4k\over(2\pi)^4}
{k^2\over(m^2_V-k^2)^2+(\varepsilon k^0)^2}\int d^2s_1d^2s_2
\exp(ikX_{12})      \nonumber\\
&&\times\left[(\dot X_1\dot X_2)(X^\prime_1X^\prime_2)
-(\dot X_1X^\prime_2)(\dot X_2X^\prime_1)\right],
\label{act1}
\end{eqnarray}
where $X_{1,2}\equiv X(\sigma_{1,2},\tau_{1,2})$. Hereafter the prime over 
$X$ denotes the derivative with respect to the parameter along the contour, 
the overdot denotes the derivative with respect to (proper)time.
Since the remaining
logarithmic divergence at coincident $s^A_1=s^A_2$ is due to our ignorance of
the Higgs field profile at the distances smaller than $m^{-1}_H$,
equivalently, at the momenta larger than $m_H$, we insert the form factor
$m^2_H/(m^2_H-k^2)$ into the integrand of Eq.~(\ref{act1}), in order to
take into account the above feature of the Higgs field. The detailed form
of the cutoff factor is irrelevant within the logarithmic accuracy 
adopted here. Using the representation
$(m^2-k^2-i\varepsilon)^{-1}=i\int_0^\infty d\alpha
\exp[i\alpha(k^2-m^2+i\varepsilon)]$, one can integrate over $k$ to obtain
the regularized action
\begin{eqnarray}
S_{\rm reg}&=&-{m^2_V\over8g^2}\int_0^\infty{d\alpha\over\alpha^2}
e^{-\varepsilon\alpha}\left(e^{-i\alpha m^2_H}-e^{-i\alpha m^2_V}\right)
\nonumber\\
&&\times\int d^2s_1d^2s_2e^{-iX^2_{12}/4\alpha}
\left[(\dot X_1\dot X_2)(X^\prime_1X^\prime_2)
-(\dot X_1X^\prime_2)(\dot X_2X^\prime_1)\right].
\label{act2}
\end{eqnarray}
Since the dominant contribution comes
from $s^A_2=s^A_1+z^A$ close to $s^A_1$, one may use the expansion
$X^2_{12}=z^Az^B\partial_AX_\mu\partial_BX^\mu$ to integrate
over $d^2s_2=d^2z$.   The result is the action in the zeroth 
order in the vortex thickness,
$S_{\rm reg}=-\varepsilon_{\rm v}\int d^2s
\sqrt{-X^{\prime2}\dot X^2+(X^\prime\dot X)^2}$,
with $\varepsilon_{\rm v}=\pi\eta^2\ln m_H/m_V$ being
the energy per unit length. This is the familiar Nambu-Goto form.
\footnote{The longitudinal correction to the
electric field strength arising due to the time component of the vector
potential was not properly taken into
account in \cite{kozhev99}. By this reason the kinetic part of
the action appeared there to be suppressed by some large factor.}
As is known, in the physical transverse gauge $\tau=X^0$,
${\bf\dot X}\cdot{\bf X}^\prime=0$, ${\bf X}^{\prime2}=1-
{\bf\dot X}^2$, the equation of motion acquires the simple form
\begin{equation}
\ddot{\bf X}-{\bf X}^{\prime\prime}=0.
\label{eqmong}
\end{equation}
However, in the situations when either the first stages of the string loop 
collapse are considered, or in the case of the long wave perturbations 
propagating on the string, or both, the transverse velocity is 
nonrelativistic. Then the retardation is inessential, and one can neglect 
$k^0$ in comparison with the large gauge boson mass $m_V$ in the propagator of 
Eq.~(\ref{fieldsk}). This permits one to integrate over $k^0$ 
to obtain the mixed, $({\bf k},t)$, Fourier representation for all necessary
quantities. It is just this case that should be kept in mind in what follows.

Let us discuss the possible modifications of Eq.~(\ref{eqmong}) by the
presence of an induced Chern-Simons (CS) term.
The nonzero density of fermions provides the naturally preferable reference
frame in which the fermionic sea is at rest. Then  all the 3D vectors
are taken relative to this reference frame.
In principle, the presence of the CS term in the action,
Eq.~(\ref{delac}), disturbs the gauge field configuration of the string,
see Ref.~\cite{bimonte95}. However, this disturbance does not alter the
form of Eq.~(\ref{eqmong}), provided the chemical potential is
sufficiently small, namely,
$\mu\ll m_V(2\pi/g)^2.$
Indeed, the field
equation for the strength of magnetic field, in the presence of the
CS term, becomes
\begin{equation}
\mbox{\boldmath$\nabla$}\times{\bf H}=m^2_V({1\over g}
\mbox{\boldmath$\nabla$}\chi-{\bf A})+{\mu g^2\over2\pi^2}{\bf H}.
\label{hstren}
\end{equation}
Hereafter the explicit dependence on time is omitted from all the notations.
Then the Fourier components of the corrections to ${\bf H}$ and
$(\mbox{\boldmath$\nabla$}\chi/g-{\bf A})$ that arise
due to a nonzero chemical potential  are found to be
\begin{eqnarray}
\delta{\bf H}({\bf k})&=&{i\mu g^2\over2\pi^2}\frac{[{\bf k}\times
{\bf H}({\bf k})]}{{\bf k}^2+m^2_V},    \nonumber\\
\delta(\mbox{\boldmath$\nabla$}\chi/g-{\bf A})({\bf k})&=&
{\mu g^2\over2\pi^2}\frac{{\bf H}({\bf k})}{{\bf k}^2+m^2_V},
\label{corr}
\end{eqnarray}
where ${\bf H}({\bf k})=2\pi g^{-1}\int d\sigma{\bf X}^\prime
\exp(-i{\bf k}\cdot{\bf X})m^2_V/({\bf k}^2+m^2_V)^{-1}$ 
is the unperturbed strength 
of  magnetic field, while $A_0+\dot\chi/g$ remains unchanged.
The leading correction is perpendicular to unperturbed fields. The
action being quadratic in the fields remains unchanged to first order in
$\mu$. Interestingly enough, but even the mirror-invariant straight string
acquires a nonzero
magnetic-like helicity in the presence of a nonzero fermionic
density characterized by $\mu\not=0$. In fact, using Eq.~(\ref{corr}), one
finds the induced helicity to be
\begin{equation}
h_{\rm induced}=\int d^3x({\bf H}\cdot\delta{\bf A}+
\delta{\bf H}\cdot{\bf A})
={\mu g^2\over\pi^2}\int{d^3k\over(2\pi)^3}
\frac{|{\bf H}({\bf k})|^2}{{\bf k}^2+m^2_V}.
\label{indhel}
\end{equation}
Since ${\bf H}({\bf k})$ for the z-directed straight string is
${\bf H}({\bf k})=(2\pi)^2g^{-1}m^2_V({\bf k}^2+m^2_V)^{-2}
\delta(k_z){\bf e}_z$, ${\bf e}_z$ being the unit vector in z direction,
the induced helicity per unit length $L$ evaluated from Eq.~(\ref{indhel})
is $h_{\rm induced}/L=\mu/2\pi$.
Its contribution to the action is
quadratic in $\mu$ and can be neglected under the condition adopted above.

The situation changes drastically when the string network is intrinsically
mirror-non-invariant.
The Chern-Simons number,
\begin{equation}
n_{\rm CS}={g^2\over8\pi^2}\int d^3x\varepsilon_{ijk}
A_iF_{jk},
\label{ncs1}
\end{equation}
is proportional to the helicity of the gauge field. In the case of the string
network it was evaluated to be $n_{\rm CS}=2\sum_{a<b}L[a,b]+\sum_{a}W[a]$
\cite{vachas94,sato95,kozhev95},
where the first term comes from the linking of any pair of the strings
$a,b$ characterized by the linking number $L[a,b]$ \cite{vachas94} while
the second one does from the writhing of each individual string $a$
\cite{sato95,kozhev95} characterized by the writhing number $W[a]$
\cite{moffatt}.
The correction to the total action is as in Eq.~(\ref{delac}), with
$n_{\rm CS}$ from Eq. (\ref{ncs1}). To obtain the equation of motion by
varying the action over ${\bf X}$, one should include the variation of the
Chern-Simons number $n_{\rm CS}$.
To this end, the Fourier representation of this number
found earlier \cite{kozhev95,kozhev99},
\begin{equation}
n_{\rm CS}=\sum_{a,b}\int d\sigma_ad\sigma_b\int d^3k
\exp\left(-i{\bf k}\cdot{\bf X}_{ab}\right)
\frac{i{\bf k}\cdot\left[{\bf X}^\prime_a(\sigma_a)
\times{\bf X}^\prime_b(\sigma_b)\right]}{{\bf k}^2}
\left(\frac{m^2_V}{{\bf k}^2+m^2_V}\right)^2,
\label{nfour}
\end{equation}
${\bf X}_{ab}\equiv{\bf X}_a(\sigma_a)-{\bf X}_b(\sigma_b)$,
is useful. When varying this expression over the string contour
${\bf X}$, one should have in mind that the terms with $a\not=b$ vanish,
because they contribute to the linking number of any pair of strings which
is the topological invariant. The typical term with $a=b$ being the
variation of the writhing number $W$ \cite{kozhev99}, after the
integration by parts \footnote{The surface terms vanish identically for the
closed string, and can be made vanishing by imposing the condition of
periodicity in the case of the open string. \label{fn1}},
and dropping the string label $a$,
becomes
\begin{eqnarray}
\delta W&=&\int d^3k\left(\frac{m^2_V}{{\bf k}^2+m^2_V}\right)^2
\int d\sigma_1d\sigma_2
\exp\left(-i{\bf k}\cdot{\bf X}_{12}\right)
\left[\delta {\bf X}(\sigma_1)-
\delta {\bf X}(\sigma_2)\right]      \nonumber\\
&&\cdot\left[{\bf X}^\prime(\sigma_1)
\times{\bf X}^\prime(\sigma_2)\right],
\label{delw}
\end{eqnarray}
where ${\bf X}_{12}\equiv{\bf X}(\sigma_1)-{\bf X}(\sigma_2)$, $\sigma_{1,2}$
refers now to the same contour. Since
$$
\int{d^3k\over(2\pi)^3}\left({m^2_V\over{\bf k}^2+m^2_V}
\right)^2\exp(-i{\bf k}\cdot{\bf X})
={m^3_V\over8\pi}\exp(-m_V|{\bf X}|),
$$
the contribution of the remote arclength segments is suppressed
exponentially with the distance between them, and one may use the expansion
${\bf X}(\sigma_2)={\bf X}(\sigma_1)
+{\bf X}^\prime(\sigma_1)(\sigma_2-\sigma_1)+\cdots$ up to the third order,
to integrate over $z=\sigma_2-\sigma_1$. One obtains
\begin{equation}
\delta W=
{m^3_V\over8\pi}\int d\sigma\delta{\bf X}\cdot[{\bf X}^\prime\times
{\bf X}^{\prime\prime\prime}]
\int_{-\infty}^\infty dz(-z^2)
\exp(-m_V|z|)
={1\over2\pi}\int d\sigma\delta{\bf X}\cdot[{\bf X}^\prime\times
{\bf X}^{\prime\prime\prime}].
\label{var}
\end{equation}
The variation of the Chern-Simons number of a single string reduces
to the variation of the writhing number (\ref{var}).
Then the equation of motion obtained upon the variation of the total
action, becomes
\begin{equation}
\ddot{{\bf X}}-{\bf X}^{\prime\prime}
-{\mu\over4\pi\varepsilon_{\rm v}}[{\bf X}^\prime\times
{\bf X}^{\prime\prime\prime}]=0.
\label{eqmo}
\end{equation}
The effect of a nonzero
fermionic number characterized by nonzero $\mu$, on the string motion
is essentially three-
dimensional and is possible only for parity-non-invariant contours.
It manifests as an additional force perpendicular to both the tangent,
${\bf X}^{\prime}$, and normal, ${\bf n}$, vectors of the string. Indeed,
using the Frenet equations,
\begin{equation}
{\bf X}^{\prime\prime}=\kappa{\bf n}\mbox{, }
{\bf n}^\prime=-\kappa{\bf X}^\prime+\tau{\bf b}\mbox{, }
{\bf b}^\prime=-\tau{\bf n},
\label{frenet}
\end{equation}
where $\kappa$, $\tau$, and  ${\bf b}$ are, respectively, the curvature, the
torsion, and the binormal vector of a curve,
one can represent Eq. (\ref{eqmo}) in the form
\begin{equation}
\ddot{{\bf X}}=\kappa
\left(1-{\tau\mu\over4\pi\varepsilon_{\rm v}}\right){\bf n}
+{\kappa^\prime\mu\over4\pi\varepsilon_{\rm v}}{\bf b}.
\label{eqmo1}
\end{equation}
The first term on the right hand side is the usual one originating from the
tension force but corrected for the interaction  with the fermionic
matter. The second term is completely new.
This term, together with the above correction to the tension, are the
classical manifestations of the purely quantum  phenomenon of an anomaly.
As is argued in Ref. \cite{rubakov85}, this is due to the appearance of new
fermionic levels from the Dirac sea upon applying an external gauge field,
in the present case, the variable gauge field of the moving string segment.
Note that the time derivative of the writhing number, $\dot W[a]=\int d\sigma
\dot{\bf X}\cdot[{\bf X}^\prime\times
{\bf X}^{\prime\prime\prime}]/2\pi$, vanishes for the translational motion
of the string \cite{kozhev99}. Hence, to first order in chemical potential,
only the dynamical internal motion of the parity-non-invariant gauge string
is sensitive to the external uniform background density of fermions.
Note the  difference of an additional force in the equation of
motion (\ref{eqmo})  with another one
obtained in Ref. \cite{uri}. An additional force besides the tension
found there depends on
the string local velocity and in this sense is anologous to the known Magnus
force acting even on the straight string. In our case, the force is of purely
chiral origin. It is independent of the local velocity, but acts only on the
strings with the mirror-non-invariant contour.

Leaving the time-dependent configurations for a future work,
let us find the static solutions to Eq. (\ref{eqmo}) and perform their
stability analysis. As is evident from Eq.~(\ref{eqmo1}), there are two
static solutions. The first solution is $\kappa=0$, $\kappa^\prime=0$ which
means the straight line, say,
${\bf X}(\sigma)=\sigma{\bf e}_z$.
The second one is
$\kappa^\prime=0\mbox{, }\tau=4\pi\varepsilon_{\rm v}/\mu$,
which means a curve with the constant
curvature $\kappa$ and torsion $\tau$. An explicit form of the contour in
this case  that can be found from solving the Frenet equations (\ref{frenet}),
is a helix,
\begin{equation}
{\bf X}(\sigma)=R\left[{\bf e}_x\cos\left({\sigma\over a}\right)+
{\bf e}_y\sin\left({\sigma\over a}\right)\right]+
{\bf e}_z{l\sigma\over2\pi a},
\label{heleq}
\end{equation}
whose radius
$R$ and step $l$ are not arbitrary but subjected to the condition
\begin{equation}
l/2\pi a^2=4\pi\varepsilon_{\rm v}/\mu.
\label{step}
\end{equation}
Hereafter we denote $a=\sqrt{R^2+(l/2\pi)^2}$. As usual, the stability
analysis demands the study of the negative modes of the second variational
derivative of the energy functional. Using the expression for $\delta W$
(\ref{delw}) to obtain the second variational derivative of this
functional in the form
\begin{equation}
{\delta^2E\over\delta X_i(\sigma)\delta X_j(\sigma^\prime)}=
-\left[2\varepsilon_{\rm v}\delta_{ij}\partial^2_\sigma
+\mu\varepsilon_{ijk}
\left(2X^{\prime\prime\prime}_k\partial_\sigma
+3X^{\prime\prime}_k\partial^2_\sigma+X^\prime_k\partial^3_\sigma
\right)/2\pi\right]
\delta(\sigma-\sigma^\prime),
\nonumber
\end{equation}
one gets, after the integration by parts (see footnote \ref{fn1}),
\begin{equation}
\delta^2E=\varepsilon_{\rm v}\int d\sigma\left(
\mbox{\boldmath$\xi$}^{\prime2}+\mu{\bf X}^\prime\cdot
[\mbox{\boldmath$\xi$}^\prime\times\mbox{\boldmath$\xi$}^{\prime\prime}]
/4\pi\right),
\label{vare}
\end{equation}
where ${\bf X}$, $\mbox{\boldmath$\xi$}$ are, respectively, the unperturbed
contour and its variation. The latter should obey the condition
\begin{equation}
\mbox{\boldmath$\xi$}^{\prime}\cdot{\bf X}^\prime=0,
\label{gcond}
\end{equation}
in order to preserve the gauge choice ${\bf X}^{\prime2}=1$ of the static
problem. First, consider the straight line
solution. Then Eq.~(\ref{gcond}) reduces
to $\xi^\prime_3=0$, and the second variation of the energy functional
becomes
\begin{eqnarray}
\delta^2E&=&\varepsilon_{\rm v}\int d\sigma\left[\xi^{\prime2}_1+
\xi^{\prime2}_2+{\mu\over4\pi\varepsilon_{\rm v}}\left(
\xi^{\prime}_1\xi^{\prime\prime}_2-\xi^{\prime}_2\xi^{\prime\prime}_1
\right)\right]   \nonumber\\
&&=\varepsilon_{\rm v}\sum^\infty_{n=1}\left(\xi^2_{1n}+\xi^2_{2n}+
{\mu n\over\varepsilon_{\rm v}L}\xi_{1n}\xi_{2n}\right)\left({2\pi n\over L}
\right)^2,
\label{straight}
\end{eqnarray}
where the use is made of an expansion of $\mbox{\boldmath$\xi$}$ into the
normal modes
\begin{equation}
\mbox{\boldmath$\xi$}(\sigma)=\sqrt{2/L}\sum_{n=1}^\infty
\mbox{\boldmath$\xi$}_n\cos(2\pi n\sigma/L),
\label{modes}
\end{equation}
found from the demand of the
periodicity in $\sigma$ with the period $L$. Diagonalisation of the expression
in the second
line in Eq.~(\ref{straight}) shows that the modes with
$n<2\varepsilon_{\rm v}L/\mu$ are stable while higher ones are unstable.
One should set $L\to\infty$ for the infinitely straight string, which means
that such a string in the fermionic matter is always stable.
However, it is known \cite{z} that Z-strings terminating on monopoles may have
the finite length. The latter can be chosen such that one may neglect the mass
of the monopoles as compared to the total energy of Z-string, in order
not to take into account the monopole contribution to the action.
Then one comes to the conclusion that sufficiently high modes on the straight
Z-string of finite length destabilize it in the presence of the fermionic
matter with the nonzero chemical potential.

The stability analysis of the helical solution
(\ref{heleq}) can be reduced to the case of the straight string by means
of the change of variables. Indeed,
$\xi^\prime_3$ found from Eq.~(\ref{gcond}) is
$$\xi^\prime_3={2\pi R\over l}\left(\xi^\prime_1\sin{\sigma\over a}
-\xi^\prime_2\cos{\sigma\over a}\right).$$
Substituting this expression into Eq.~(\ref{vare}) and using Eq.~(\ref{step}),
one obtains
\begin{eqnarray}
\delta^2E&=&\varepsilon_{\rm v}\int d\sigma\left[\xi^{\prime2}_1+
\xi^{\prime2}_2+\left({2\pi R\over l}\right)^2\left(
\xi^{\prime}_1\sin{\sigma\over a}
-\xi^{\prime}_2\cos{\sigma\over a}
\right)^2+\left({2\pi a\over l}\right)^2a
\left(\xi^\prime_1\xi^{\prime\prime}_2-\xi^{\prime}_2\xi^{\prime\prime}_1
\right)\right.    \nonumber\\
&&\left.-\left({2\pi R\over l}\right)^2
\left(\xi^{\prime}_1\cos{\sigma\over a}+\xi^{\prime}_2\sin{\sigma\over a}
\right)^2\right].
\label{delehe}
\end{eqnarray}
Now, introducing new variables $\Xi_{1,2}$ according to the relation
$$\left(\xi^\prime_1\atop\xi^\prime_2\right)=
{1\over\sqrt{2}}\pmatrix{\cos{\sigma\over a}&-\sin{\sigma\over a}\cr
\sin{\sigma\over a}&\cos{\sigma\over a}\cr}
\left(\Xi^\prime_1\atop{l\over2\pi a}\Xi^\prime_2\right),
$$ one finds from Eq.~(\ref{delehe}) that
\begin{eqnarray}
\delta^2E&=&\varepsilon_{\rm v}\int d\sigma\left[\Xi^{\prime2}_1+
\Xi^{\prime2}_2+{\pi a^2\over l}\left(
\Xi^{\prime}_1\Xi^{\prime\prime}_2-\Xi^{\prime}_2\Xi^{\prime\prime}_1
\right)\right]          \nonumber\\
&&=\varepsilon_{\rm v}\sum^\infty_{n=1}\left(\Xi^2_{1n}+\Xi^2_{2n}+
{(2\pi a)^2n\over lL}\Xi_{1n}\Xi_{2n}\right)
\left({2\pi n\over L}
\right)^2,
\label{helix}
\end{eqnarray}
where the expansion into the  normal modes analogous to Eq.~(\ref{modes})
in the case of the straight string is used.
One can see, using Eq.~(\ref{step}), that the stability condition for the
helical solution coincides with that in the case of the straight string:
the modes with $n<lL/2\pi^2a^2=2\varepsilon_{\rm v}L/\mu$ are stable.
Again, an infinitely long helix is stable, while the higher modes on the
helix of finite length become negative in the presence of the fermionic
matter with $\mu\not=0$.

To conclude, two new effects due to the presence of fermionic matter with
finite density important for the gauge string configurations are found.
First, a nonzero magnetic-like helicity is induced on such a non-chiral
object as the straight string. Second, the fermionic
background  exerts an influence on the motion of the
string with intrinsically parity-non-invariant contours as an additional
force besides the tension.
The correction to the string equation of motion found here should be compared
with the lowest order finite thickness correction, $\sim(\kappa/m_H)^2$,
with $\kappa$ being the curvature of the contour, 
calculated in \cite{arodz,bonjour}. Taking, for instance, the case of the
helical contour of the step $l$ one can find the condition of the dominance of
the $\mu\not=0$ correction in the form $\mu l\gg\lambda^{-2}\ln\lambda$,
where the self-interaction coupling of the Higgs field is large,
$\lambda^2\gg1$, in the London limit adopted here.
Static solutions corresponding to the balance
of the above forces are found, and their stability is explored. The stability
refers to that in the background of fermions. If one has in mind the
electroweak Z-string \cite{z} that definitely contributes to the anomaly
of the sum of the baryon and lepton currents, it possesses its own
instabilities, see recent review \cite{achuc}. The question of whether these
instabilities can be neutralized or not is still an open one \cite{achuc}.
Despite this, the Abelian-like spontaneously broken gauge model of the vector
field coupled to the chiral fermions considered in the present work can be
viewed as an ingredient of some wider theoretical framework and by this
reason might be of interest as an additional demonstration of how
purely quantum phenomenon of an anomaly manifests on the macroscopic
scale.

\end{document}